      [1999/12/01 v1.4c Il Nuovo Cimento]
\documentclass{cimento}

\usepackage{graphicx}  

\title{MITSuME---Multicolor Imaging Telescopes for Survey and Monstrous
Explosions}
\shorttitle{MITSuME}
\author{T.~Kotani\from{tokyotech}\thanks{TK is supported by a 21st Century COE Program at
Tokyo Tech ``Nanometer-Scale Quantum Physics'' by the
Ministry of Education, Culture, Sports, Science and Technology.}\ETC,
 N.~Kawai\from{tokyotech},
 K.~Yanagisawa\from{oao},
 J.~Watanabe\from{naoj},
 M.~Arimoto\from{tokyotech},
 H.~Fukushima\from{naoj},
 T.~Hattori\from{oao},
 M.~Inata\from{oao},
 H.~Izumiura\from{oao},
 J.~Kataoka\from{tokyotech},
 H.~Koyano\from{oao},
 K.~Kubota\from{tokyotech},
 D.~Kuroda\from{sokendai},
 M.~Mori\from{icrr},
 S.~Nagayama\from{oao},
 K.~Ohta\from{kyoto},
 T.~Okada\from{oao},
 K.~Okita\from{oao},
 R.~Sato\from{tokyotech},
 Y.~Serino\from{tokyotech},
 Y.~Shimizu\from{oao},
 T.~Shimokawabe\from{tokyotech},
 M.~Suzuki\from{tokyotech},
 H.~Toda\from{uair},
 T.~Ushiyama,
 Y.~Yatsu\from{tokyotech},
 A.~Yoshida\from{aoyama},
        \atque
 M.~Yoshida\from{oao}
}

\instlist{
\inst{tokyotech} Tokyo Tech - O-okayama, Tokyo, Japan
\inst{oao} Okayama Astrophysical Observatory/NAOJ - Kamogata, Okayama, Japan
\inst{naoj} National Astronomical Observatory Japan -  Osawa, Tokyo, Japan
\inst{sokendai} Sokendai - Yoshinodai, Kanagawa, Japan
\inst{icrr} Institute of Cosmic-Ray Research - Kashiwanoha, Chiba, Japan
\inst{kyoto} Kyoto University - Sakyo, Kyoto, Japan
\inst{uair} University of the Air - Mihama, Chiba, Japan
\inst{aoyama} Aoyama Gakuin University - Fuchinobe, Kanagawa, Japan}

\PACSes{\PACSit{00.00}{}
\PACSit{---.---}{\ldots}}

\begin{document}

\newcommand{\oao}{{\it OAO 50 cm}}
\newcommand{\oaowf}{{\it OAOWFC}}
\newcommand{\akeno}{{\it Akeno 50 cm}}
\newcommand{\tri}{{\it Tricolor Camera}}

\maketitle

\begin{abstract}
Development of MITSuME is reported.  Two 50-cm optical telescopes have
been built at Akeno in Yamanashi prefecture and at Okayama Astrophysical
Observatory (OAO) in Okayama prefecture.  Three CCD cameras for
simultaneous $g'R_CI_C$ photometry are to be mounted on each focal
plane, covering a wide FOV of about $30" \times 30"$.  The limiting
magnitude at $V$ is fainter than 18.  In addition to these two optical
telescopes, a 91-cm IR telescope with a $1^\circ \times 1^\circ$ field
of view is being built at OAO, which performs photometry in $YJHK$
bands.  These robotic telescopes can start the observation of
counterparts of a GRB within a minute from an alert.  We aim to obtain
photometric redshifts exceeding 10 with these telescopes.  The
performance and the current construction status of the telescopes are
presented.
\end{abstract}

\section{The MITSuME Project}
Three IR/optical robotic telescopes for prompt observation of GRBs and
afterglows are being developed.  The project, {\it MITSuME,} is promoted
by Tokyo Tech, National Astronomical Observatory of Japan (NAOJ), the
Institute of Cosmic-Ray Research (ICCR), Kyoto University, and Aoyama
Gakuin University.  {\it MITSuME} stands for {\it Multicolor Imaging
Telescopes for Survey and Monstrous Explosions,} and also means {\it
three eyes} in Japanese.  The objective of the project is multi-band
photometry from $K_s$ to $g'$ of GRBs and afterglows within tens of
seconds and detection/determination of cosmological events with
redshifts exceeding 10.  Two optical telescopes, \akeno\/ and \oao, have
been constructed at Akeno Observatory of the ICCR, Akeno in Yamanashi
prefecture, and at Okayama Astrophysical Observatory (OAO) of NAOJ,
Kamogata in Okayama prefecture.  Each telescope has a \tri\/ capable of
$g'R_CI_C$-bands photometry.  An existing 91-cm telescope at OAO is
being converted to an IR telescope, \oaowf\/ (OAO Wide Field Camera). It
is designed to have a wide field of view of $56'\times56'$ and perform
$YJHK_S$ photometry~\cite{yanagisawa02}.  These three telescopes are to be
automatically operated and respond to GRB alerts.

\section{Telescopes}
The specification of each telescope is shown in Table~\ref{tab:spec}.
\oao\/ can be maneuvered at a speed of 4 deg s$^{-1}$, or within 45 s to
any direction, and \akeno\/ at a speed of 9 deg s$^{-1}$, within 20 s to
any direction.  \oao\/ and \akeno\/ have large field of views of
$30'\times30'$ and are suitable for a source with a position uncertainty
up to $15'$, which is typical for an on-board localization by GRB
monitoring missions.  \oaowf\/ has an even larger field of view of
$1^\circ\times1^\circ$, and will be used for a survey of Mira variables
when it is not occupied in a follow-up observation.  A mixture system of
$g'$, $R_C$, and $I_C$ is adopted for \akeno\/ and \oao\/ instead of the
standard Johnson-Cousins system~\cite{johnson65,cousins78} or the SDSS
system~\cite{smith02}.  We have chosen $g'$ rather than $V$, because the
broader bandwidth of the former would give a better
sensitivity~\cite{smith02}.  We have chosen $I_C$ rather than $i'$ for
the same reason.  By selecting $R_C$ instead of $r'$ we can avoid
artificial lights in the night sky.  For \oaowf, a $YJHK_S$ system is
adopted~\cite{yanagisawa02,hillenbrand02}.

\begin{table}
  \caption{Specification.}
  \label{tab:spec}
\footnotesize
  \begin{tabular}{lccc}
    \hline
	                  &\oaowf	        &\oao	        &\akeno\\
\hline
Maneuverability	          &1.5 deg s$^{-1}$	&4 deg s$^{-1}$ & 9 deg s$^{-1}$ \\
Mirror Diameter           &910 mm               &500 mm         &500 mm\\
Focal Length	          &2260 mm              &3250 mm	&3000 mm\\
F Number	          &2.5	                &6.5	        &6.0\\
FOV	                  &$56'\times56'$	&$26'\times26'$ &$28'\times28'$\\
Bands	                  &$YJHK_S$	        &$g'R_CI_C$	&$g'R_CI_C$\\
Limiting Mag.\	          &$J=16.6$             &$V=18.4$	&$V=18.2$\\
(10 min, S/N=10)          &$H=15.8$             &$R=18.5$	&$R=18.2$\\
	                  &$K=15.4$             &$I=17.7$	&$I=17.5$\\
Location                  &\multicolumn{2}{c}{OAO/NAOJ}               &Akeno Observatory/ICRR\\
   &\multicolumn{2}{c}{Kamogata, Okayama, Japan} &Akeno, Yamanashi, Japan\\
                          &\multicolumn{2}{c}{$133^\circ35'$E, $34^\circ34'$N, 372 m
A. S. L.}    &$138^\circ30'$E, $35^\circ47'$N, 900m A. S. L.\\
    \hline
  \end{tabular}
\end{table}

\begin{figure}
\begin{minipage}{0.5\textwidth}
\resizebox{1\textwidth}{!}{\includegraphics{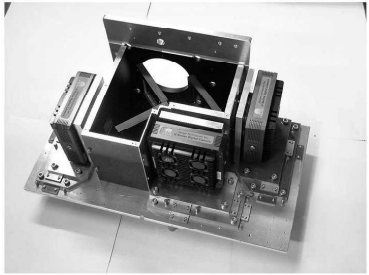}}
\end{minipage}
\begin{minipage}{0.49\textwidth}
\caption{Top View of {\it Tricolor Camera}}\label{fig:tricolor} A top
panel is removed.  The light led in the camera from above is divided
into three bands with two dichroic mirrors and a gold-coated mirror and
distributed to three {\it Alta U6} CCDs
\end{minipage}
\end{figure}

A \tri\/ is to be mounted on each focal plane of \oao\/ and \akeno.  A
prototype of \tri\/ is shown in Fig.~\ref{fig:tricolor}.  The \tri\/
employs three {\it Alta U-6}\/ cameras (Apogee Instruments Inc.), and
each {\it Alta U-6}\/ has a {\it KAF-1001E}\/ CCD (Kodak) with
$1024\times1024$ pixels.  The pixel size is $24 \mu$m $\times 24 \mu$m,
or $1.6"\times1.6"$ at the focal planes.  Three images of different
bands are simultaneously taken.

As for \oaowf, an array of {\it HAWAII-2 RG PACE}\/ (Rockwell Science)
is employed as a focal plane detector, and its housing is being
developed.

\section{Robotic System}
\begin{figure}
\begin{minipage}{0.66\textwidth}
\resizebox{1\textwidth}{!}{\includegraphics{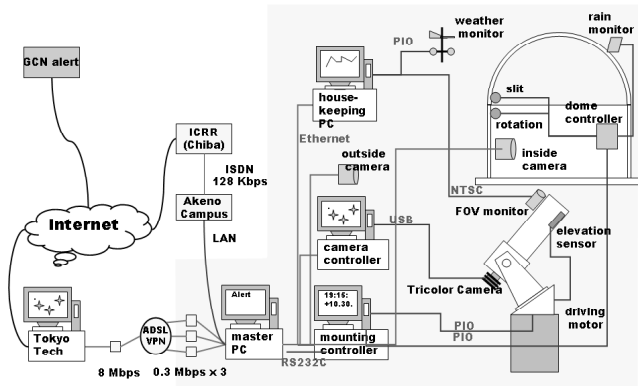}}
 \end{minipage}
\begin{minipage}{0.33\textwidth}
\caption{Robotic System to control \akeno}\label{fig:robotic}
The system resides in a 4-m dome at Akeno (shaded area).  It is
 connected with Tokyo Tech via ADSLs.  
\end{minipage}
\end{figure}

The robotic system to control \akeno\/ is shown in
Fig.~\ref{fig:robotic}.  When the mastering PC receives an alert, it
points the telescope to the GRB location or waits until the target
becomes observable.  CCD images are transfered to Tokyo Tech.  An
automated search for an optical counterpart is to be performed.  The
system is designed to function without on-site maintenance for weeks.

\section{Current Status}
\paragraph{\oaowf}  Conversion from an existing
telescope is in progress.  The designing of a new optics, detector, and
cooling system has been done.  A first light is scheduled in 2005.

\begin{figure}
\begin{minipage}{0.75\textwidth}
 \resizebox{1\textwidth}{!}{\includegraphics{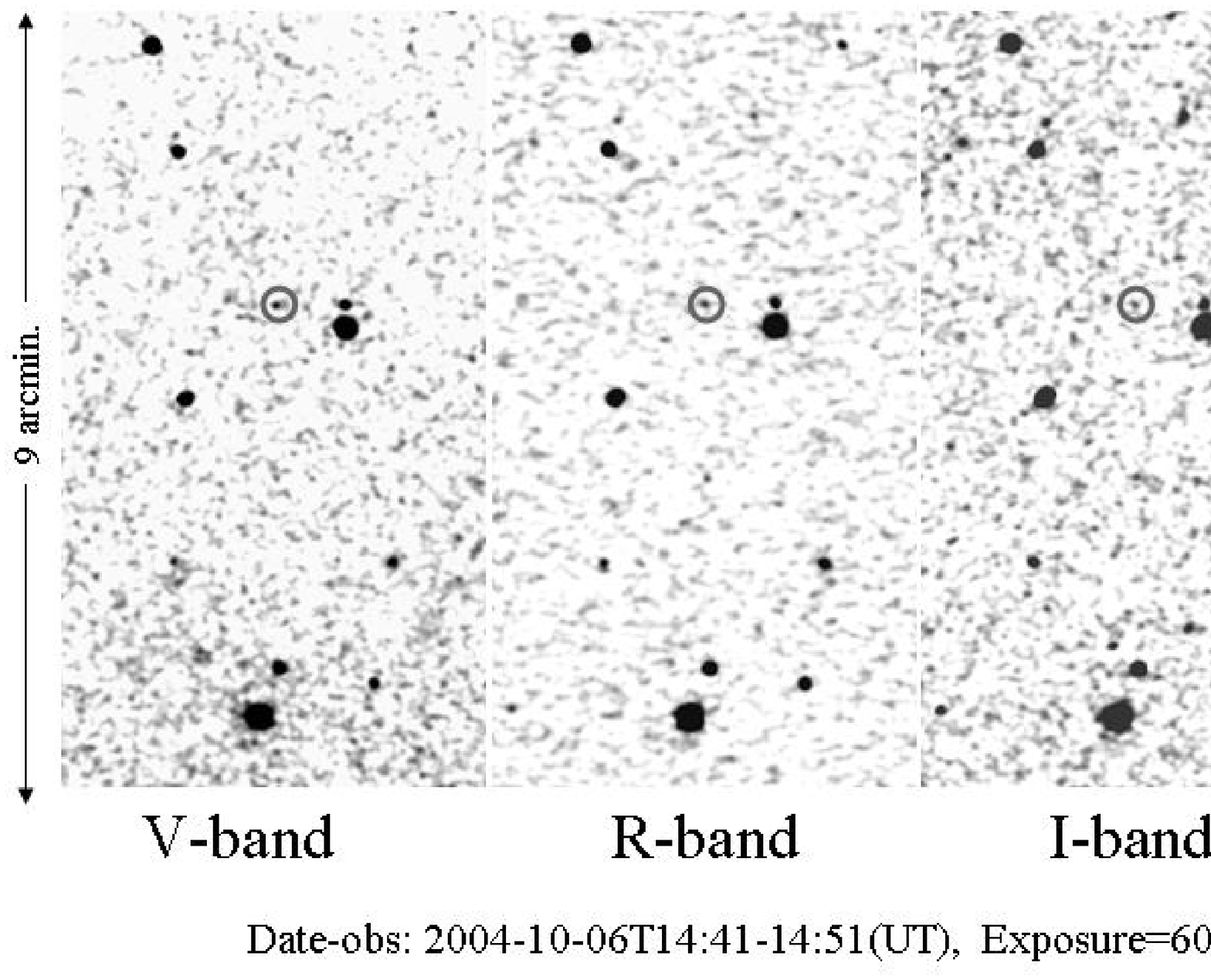}}
\end{minipage}
\begin{minipage}{0.24\textwidth}
\caption{GRB041016}\label{fig:grb} $V$, $R$, and $I$ images of GRB041016
taken with \oao\/ and the DSS image of the same field of view.  The
position of GRB041016 is indicated with a circle.
\end{minipage}
\end{figure}

\paragraph{\oao}  The construction of the telescope and a 4-m dome has
been done.  Calibration and performance verification are almost done. It
is currently operated by a human operator.  The images of GRB041016
taken with the prototype \tri\/ is shown in
Fig.~\ref{fig:grb}~\cite{kuroda04}.  The afterglow is successfully
detected, which is not recognized in the DSS image.

\paragraph{\akeno}  The construction has been done.  Calibration and
performance verification are almost done.  A robotic program is being
developed based on RIBOTS' program~\cite{kohama01,urata01}.  It is
operated by a human operator at a remote site.  GRB050124 and GRB050209
were observed~\cite{yatsu0501,yatsu0502}.

\begin{thebibliography}{99}
 \bibitem{yanagisawa02} \BY{{Yanagisawa} K., \etal}
	in \TITLE{The Proceedings of the IAU 8th Asian-Pacific Regional
	Meeting, Volume II},
	edited by \NAME{Ikeuchi S., Hearnshaw J., \atque Hanawa T.}
	(Astronomical Society of Japan, Tokyo) 2002, p.~83.

\bibitem{johnson65} \BY{Johnson H. L.}
  \IN{Astron.\ J.}{141}{1965}{923}.

\bibitem{cousins78} \BY{Cousins A. W. J.}
  \IN{Mon.\ Not.\ Astron.\ Soc.\ South Africa}{37}{1978}{8}.

\bibitem{smith02} \BY{Smith J. A., \etal}
  \IN{Astron.\ J.}{123}{2002}{2121}.

\bibitem{hillenbrand02} \BY{Hillenbrand  L. A., \etal}
	\IN{Publ.\ Astron.\ Soc.\ Pacific}{114}{2002}{708}.

\bibitem{kuroda04} \BY{Kuroda D., Yanagisawa K., \atque Kawai N.}
  \IN{GCN}{2818}{2004}{}.

\bibitem{kohama01} \BY{Kohama M., \etal}
   in \TITLE{New Century of X-ray Astronomy},
                   edited by \NAME{Inoue  H.  \atque Kunieda H.}
                   (Astronomical Society of the Pacific, San Francisco) 2001, p.~558.

\bibitem{urata01} \BY{Urata Y., \etal}
   in \TITLE{Small Telescope Astronomy on Global Scales},
                   edited by \NAME{Paczynski B., Chen W. P., \atque
	Lemme C.}
                   (Astronomical Society of the Pacific, San Francisco) 2001, p.~155.

\bibitem{yatsu0501} \BY{Yatsu Y. \atque Kawai N.}
  \IN{GCN}{2979}{2005}{}.

\bibitem{yatsu0502} \BY{Yatsu Y., Arimoto M., \atque Kawai N.}
  \IN{GCN}{3016}{2005}{}.
\end{thebibliography}
\end{document}